\documentclass[12pt]{elsarticle}
\usepackage[utf8]{inputenc}
\usepackage[T1]{fontenc}
\usepackage{amssymb}
\usepackage{amsmath}
\usepackage[utf8]{inputenc}
\usepackage{tikz}
\usetikzlibrary{arrows.meta}
\usepackage{amsmath,amssymb,amsfonts,graphics,graphicx,dcolumn,bm,enumerate}
\usepackage{comment,natbib,appendix}
\usepackage{multirow,color}
\usepackage{chngpage}
\usepackage{afterpage}
\usepackage{xcolor}
\usepackage{amsthm}
\usepackage{natbib}
\usepackage{hyperref}
\usepackage[margin=0.8in]{geometry}
\usepackage{epstopdf}
\usepackage{float}
\usepackage{soul}

\journal{Physica A}

\begin{document}

\begin{frontmatter}

\title{Exploring the impact of multi-agent wealth exchange model on inequality reduction}

\author[inst1]{Suchismita Banerjee}\ead{suchib@bose.res.in}

\affiliation[inst1]{organization={Department of Physics of Complex systems,\\ S. N. Bose National Centre for Basic Sciences},
            city={Kolkata},
            postcode={700106}, 
            country={India}}

\begin{abstract}
Binary kinetic-exchange models, where money is shuffled between two agents at a time, reproduce the Boltzmann–Gibbs exponential wealth distribution but cannot address the multi-party trades common in real markets. We generalize the exchange rule to simultaneous interactions among more than two agents in a closed economical system. We observe, as number of agents grow, the stationary wealth distribution evolves smoothly from an exponential to an almost uniform distribution. Inequality metrics (Gini and $k$-index) has been found to fall monotonically with the increase in agents number. Compared with binary models that rely on saving propensities, which is also known to reduce inequality, we find the multi-agent interaction show a completely different behavior of inequality reduction. 
\end{abstract}


\end{frontmatter}

\section{Introduction}
\label{sec:intro}
Kinetic exchange models of money emerged as a striking analogue to classical gas kinetics, treating individual wealth holdings as conserved `energies' exchanged stochastically between economic agents. 
In the seminal work of Dragulescu and Yakovenko~\cite{dragulescu_2000,druagulescu_2001,dragulescu_2002} (see also~\cite{BM_2000}), a pair of agents is chosen at random, they pool their money, and then share it according to a uniformly distributed random fraction; at stationarity this binary collision rule yields the Boltzmann--Gibbs
exponential wealth distribution, with an effective temperature equal to the average money per agent.
Owing to its analytical tractability, this minimal, conserved money model became a benchmark for the statistical mechanics of markets, and was situated within the broader econophysics program by canonical reviews and monographs~\cite{Yako_2009,Wealth}.
 
Over the past two decades, a rich variety of extensions has been explored to capture more realistic features of economic interactions.  
Introducing a fixed or heterogeneous saving propensity among agents or multiplicative noise leads to the exponential bulk and can generate Gamma like cores with Pareto tails in the wealth distribution~\cite{chakraborti_2000,chakraborti_2002,chatterjee_2003,chatterjee_2004,chatterjee_2005,patriarca_2004,patriarca_2006,patriarca_2013,chatterjee_2007,chatterjee1_2007,chakrabarti_2009,chatterjee_2009,chakraborti_2011,basu_2008,banerjee_2006}.
Closely related exchange and aggregation formalisms, both early and modern, supply additional baselines and heuristics for heavy tails~\cite{ispolatov_1998,repetowicz_2005,Manna_2005}.

A complementary line derives these models as Boltzmann type equations and quasi invariant limits, enabling hydrodynamic closures and rigorous statements about approach to equilibrium and tail formation~\cite{cordier_2005,during_2008,toscani_2013,pareschi_2014,lim_2020}. Recent applications and variants further explore robustness and modeling choices~\cite{Bijin_2022,bijin,mohanty_2006}.
Departures from strict conservation via debt bounds, leakage or injection, or institutional frictions, alter the invariant law and can preserve or sharpen Pareto tails relative to closed settings~\cite{silva_2004,bennati_1988,Debt}. Beyond saving based mechanisms, position exchange dynamics provide a clean protocol that toggles asymptotics (exponential vs power law) once openness and trapping are admitted~\cite{PosExchange}.
Broader perspectives on self-organization and heavy tailed phenomena consolidate these patterns across complex systems~\cite{sornette_2014}. Policy augmented models examined how taxation and redistribution influence steady states and inequality, offering interpretable levers atop kinetic exchange~\cite{kulp_2019,DINIZ_2012}.
Numerous other models to simulate the asset exchange has been derived using a different interaction rules such as
Yard sale model~\cite{chakraborti_2002,YS1,YS2}, immediate exchange model~\cite{pat_2014}, slanina model~\cite{slanina_2004} to name a few. 
These models have been invoked to address the realistic wealth distribution curves and also been attempted to see their dynamics against realistic policies such as Tax redistribution~\cite{YS3}, preferential interactions~\cite{YS_net1,YS_net2}.

Despite the breadth of extensions, most baselines remain this binary exchange assumption. Yet many real world scenarios where more than two parties are involved in a transaction like auctions, clearing or netting events, joint projects are effectively simultaneous $n$–party exchanges. 
As a result, there is still no clear idea of how simultaneous $n$-agent interactions alone reshape steady state wealth. 
No systematic study has mapped the transition from the classical exponential $(n = 2)$ to the distributions produced when $n>2$ agents share wealth in a single collision. 
Consequently, the field lacks a baseline for `pure' multi-agent exchange against which hybrid or policy augmented models can be compared. 
In this paper, we fill that gap by formulating and simulating a family of pure $n$‐agent kinetic exchange rules.

From an economic standpoint, binary kinetic exchange rules can be viewed as idealized representations of anonymous, pairwise trades in competitive markets, which naturally lead to Boltzmann-Gibbs like wealth distributions under reversible dynamics~\cite{dragulescu_2000,chatterjee1_2007,patriarca_2010}. However, many real-world interactions are inherently multi-agent and institutionally asymmetric, involving small groups of firms, households or intermediaries (e.g., oligopolistic competition, joint ventures, platform mediated trades) rather than isolated bilateral encounters~\cite{cournot_1838,tirole_1988}. In this work we therefore focus on explicit $n$-agent exchange rules that, while conserving total wealth, deliberately break microscopic reversibility, in order to explore how such group level asymmetries can generate non-Boltzmann stationary wealth distributions and more unequal outcomes.


This paper is organized as follows: Section~\ref{sec:2} formalizes the $n$‐agent collision rule extending the binary limit.  
Section~\ref{sec:3} describes our choice of simulation parameters along with tests showing the robustness of the interaction mentioned in the earlier section. 
In Section~\ref{sec:4} we present and discuss the results from our simulations of multi-agent interactions and their implications on wealth inequality. 
Lastly, in section~\ref{sec:5} we summarize and conclude our findings. 

\section{Models of multi-agent kinetic wealth exchange}\label{sec:2}

\subsection{Model}
We consider a closed economy of N agents, indexed by $i=1,\ldots,N$ whose money holdings at (discrete) time $t$ are $w_i(t)\geq 0$. 
Total wealth: $W=\sum_{i=1}^{N}w_i(t)$ is conserved by construction.
At every elementary collision a subset $G_n\subset \{1,\ldots,N\}$ of fixed size $n$ is sampled uniformly without replacement; the $n$ agents instantaneously reshuffle their combined wealth while all others remain unchanged. For a given group $G_n=\{i_1,\ldots,i_n\}$ we denote $\mathcal{W}_n=\sum_{j\in G_n} w_j(t)$ and draw a vector of partition coefficients $\pi(n)=(\pi_{i_1},\ldots,\pi_{i_n})$, $\pi_{i_l}\geq 0$,$\sum_{l=1}^n \pi_{i_l}=1$. The post-interaction wealths are then $w_{i_l}(t+1)=\pi_{i_l}\, \mathcal{W}_{n}$, $\forall \,\,l=1,\ldots,n$.

The essence of any kinetic-exchange rule is therefore the probabilistic prescription for $\pi^{(n)}$.
Below we specify that prescription for $n=2,3,4,5,6$, always ensuring:
\begin{itemize}
    \item \textbf{Conservation:} $\sum_{l} \pi_{i_l}=1 \implies \sum_{i} w_{i}$ is invariant;

    \item \textbf{Symmetry:} each agent’s post-trade money depends on one random variate and on $\mathcal{W}_{n}$ only, so all participants are statistically equivalent;

    \item \textbf{Uniform marginals:} every $\pi_{i_l}$ is marginally $Unif(0,1)$.
\end{itemize}

\subsection{Binary benchmark $(n=2)$}\label{sec:2.2}
For completeness we recall the Dragulescu–Yakovenko (DY) rule \cite{dragulescu_2000}. Drawing a single uniform deviate $\epsilon \sim Unif(0,1)$ and defining $\pi_{1}=\epsilon, \pi_{2}=1-\epsilon$,
\begin{equation}
    \begin{aligned}
        w_{1}^{*}=\pi_{1}(w_{1}+w_{2}) \\
        w_{2}^{*}=\pi_{2}(w_{1}+w_{2})
    \end{aligned}
    \label{eq:k1}
\end{equation}
produces the well-known exponential steady state.

We have extended this DY model of wealth interaction to more than two agents. 
For this, we provide here the three-agent interaction as one of the bases for constructing multi-agent modeling. 
The other basis is the DY model, which is valid for 2 agents. 
All other multi-agent interactions can be resolved in terms of such dyadic (DY) and triadic pairs. 
For example, a 4-agent system would have the total wealth divided between two pairs of agents, and each pair would individually perform kinetic exchange to distribute the wealth among themselves. 
A five-agent system would have the total wealth distributed among a pair of 3 and 2 agents. Subsequently, the triad would distribute the wealth following the triadic interaction, and the other two would distribute the wealth among themselves via DY interaction. 
Below, we describe the triadic interaction and also how the combination of triadic and DY interaction can be used to build the 4-agent tetradic, 5-agent pentadic, and 6-agent hexadic systems. 

\subsection{Triadic exchange $(n=3)$}\label{sec:2.3}
Following the conservation of wealth in each interaction event, we need three uniform random numbers that sum to unity. 
For such, a convenient construction would be the following, 
\begin{equation}
    \begin{aligned}
        \epsilon_{1}\sim Unif[0,1] \\
        \epsilon_{2}=
        \begin{cases}
    \epsilon_{1}-\frac{1}{2},& \text{if } \epsilon_{1}\geq \frac{1}{2}\\
    \epsilon_{1}+\frac{1}{2},              & \text{if } \epsilon_{1}< \frac{1}{2}
\end{cases} \\
        \epsilon_{3} = \frac{3}{2}-(\epsilon_{1}+\epsilon_{2})
    \end{aligned}
    \label{eq:k2}
\end{equation}
One readily checks $\epsilon_{1,2,3}\in (0,1)$ and $\epsilon_{1}+\epsilon_{2}+\epsilon_{3}=\frac{3}{2}$.
Setting $\pi_{l}=\frac{2\,\epsilon_{l}}{3}\,,\,\,l=1,2,3$, enforces $\sum \pi_{l}=1$.
The update rule is therefore
\begin{equation}
w_{i}^{*}=\frac{2\epsilon_i}{3}\,(w_{1}+w_{2}+w_{3}),\quad i=1,2,3
\label{eq:k3}
\end{equation}
A three-agent interaction in our model is driven by a single independent random seed $\epsilon_1 \sim \mathrm{Unif}(0,1)$, from which we deterministically construct $(\epsilon_1,\epsilon_2,\epsilon_3)$ via Eq.~\eqref{eq:k2}, equivalently $\epsilon_2 = (\epsilon_1 + \tfrac12)\bmod 1$ and $\epsilon_3 = \tfrac32 - (\epsilon_1 + \epsilon_2)$, so that $0<\epsilon_i<1$ and $\epsilon_1+\epsilon_2+\epsilon_3 = \tfrac32$. Defining $\pi_i = \tfrac{2}{3}\epsilon_i$ gives $0<\pi_i<\tfrac{2}{3}$ and $\sum_{i=1}^3 \pi_i = 1$, and the post-interaction wealths are $w_i^* = \pi_i (w_1+w_2+w_3)$. Although $(\epsilon_1,\epsilon_2,\epsilon_3)$ (and hence $(\pi_1,\pi_2,\pi_3)$) are correlated, the interaction is genuinely stochastic: each triad has a continuum of possible outcomes parametrized by $\epsilon_1$, which is resampled independently at every event. The piecewise-linear constraint between $\epsilon_1$ and $\epsilon_2$ is a deliberate minimal choice: it ensures exact wealth conservation, uses only linear operations on a single seed, is permutation-symmetric across agents, and yields identical uniform marginal `reshuffling' noise $\pi_i \sim \mathrm{Unif}(0,2/3)$ with $\mathbb{E}[\pi_i]=1/3$ for all $i$. Note that independent uniform marginals on $[0,1]$ would violate conservation, since $\mathbb{E}[\sum_i \pi_i]=3/2\neq 1$; restricting the support to $[0,2/3]$ is therefore required if one insists on uniform marginals and exact conservation. Our goal here is not to represent the most general three-agent trade, but to provide a controlled, analytically transparent baseline where changes in the stationary statistics can be attributed to the arity of the interaction under pooling, rather than to a change of the underlying noise law.

In summery, the interaction might seem like only two-thirds of an individual's wealth is at stake, and the rest is not taking part in the interaction.
Which is why later we show the comparison of the multi-agent wealth interaction with the two-agent kinetic exchange with saving propensity, which is solely based on the idea of keeping a portion of wealth in hand and interacting with the rest.


\subsection{Tetradic exchange $(n=4)$}\label{sec:2.4}
As discussed earlier, for 4 agents, we split the agents into two independent pairs, each carrying half of the total wealth:
\begin{equation}
    \begin{aligned}
        \epsilon_{1}\sim Unif[0,1],\,\,\,\,\,\pi_{1}=\frac{\epsilon_{1}}{2},\,\,\,\,\,\pi_{2}=\frac{1-\epsilon_{1}}{2}, \\    \epsilon_{2}\sim Unif[0,1],\,\,\,\,\,\pi_{3}=\frac{\epsilon_{2}}{2},\,\,\,\,\,\pi_{4}=\frac{1-\epsilon_{2}}{2}. 
    \end{aligned}
    \label{eq:k4}
\end{equation}
Then $\sum_{i=1}^{4} \pi_{i}=1$ and each $\pi_{i}$ is uniform. 
The $2$ arising in the denominator depicts the fact that each of the pairs has exactly half of the total wealth to distribute among themselves.
Following this, the post-trade wealth reads
\begin{equation}
\begin{aligned}
w_{1}^{*}=\dfrac{\epsilon_1}{2}\,\mathcal{W}_4,\,\,\,
w_{2}^{*}=\dfrac{1-\epsilon_1}{2}\,\mathcal{W}_4,\\
w_{3}^{*}=\dfrac{\epsilon_2}{2}\,\mathcal{W}_4,\,\,\,
w_{4}^{*}=\dfrac{1-\epsilon_2}{2}\,\mathcal{W}_4,
\end{aligned}
\label{eq:k5}
\end{equation}
with $\mathcal{W}_4=w_1+w_2+w_3+w_4$.

\subsection{Pentadic exchange $(n=5)$}\label{sec:2.5}
For 5 agents, we partition the system into a 3-agents and 2-agents pair, and realize that for 3-agents $\sum_{i=1}^3\epsilon_{i}=3/2$ and for the binary DY model $\sum_{i=1}^2\epsilon_{i}=1$.
Therefore, for 5-agents that finally leads to $\sum_{i=1}^5\epsilon_{i}=5/2$. 
Such a non-unitary sum demands the random numbers to be renormalized leading to define $\pi_{i}=\frac{2}{5}{\epsilon_{i}}$ in the following equation
\begin{equation}
w_{i}^{*}=\pi_{i}\,\mathcal{W}_5,\quad i=1,\dots ,5
\label{eq:k7}
\end{equation}
where $\mathcal{W}_5=w_1+\dots +w_5$.

\subsection{Hexadic exchange $(n=6)$}\label{sec:2.6}
For the 6-agent exchange, we consider two different approaches, one partitioning the 6 agents pool in a two independent 3-agent system and distribute the wealth according section~\ref{sec:2.3}, the other considers partitioning the pool into three binary pairs where each pair exercises DY-like kinetic exchange. 
\subsubsection{Type I (two independent triplets)}
Generate two independent triads of coefficients exactly as in section~\ref{sec:2.3}:
\begin{equation}
\{\epsilon_{1},\epsilon_2,\epsilon_3\},\,\,\,\{\epsilon_4,\epsilon_5,\epsilon_6\},\,\,\,\sum_{m=1}^{3}\epsilon_m=\frac{3}{2},\,\,\,\sum_{m=4}^{6}\epsilon_{m}=\frac{3}{2}.
\label{eq:k8}
\end{equation}
Set $\pi_{i}=\frac{\epsilon_{i}}{3} \quad i=1,\dots ,6$, so that $\sum_{i}\pi_{i}=1$.
The update rule is
\begin{equation}
w_{i}^{*}=\frac{\epsilon_i}{3}\,\mathcal{W}_6,\quad i=1,\dots ,6
\label{eq:k9}
\end{equation}
with $\mathcal{W}_6=\sum_{j=1}^{6}w_{j}$.

\subsubsection{Type II (three independent binary splits)}
Form three independent dyads as in section~\ref{sec:2.4}, each receiving one-third of the total wealth:

$\epsilon_1,\epsilon_2,\epsilon_3\sim Unif(0,1)$,
\begin{equation}
    \begin{aligned}    
     \pi_{1,2}=\frac{\epsilon_1,\,1-\epsilon_1}{3},\,\,\pi_{3,4}=\frac{\epsilon_2,\,1-\epsilon_2}{3},\,\,\pi_{5,6}=\frac{\epsilon_3,\,1-\epsilon_3}{3} .  
    \end{aligned}
    \label{eq:k10}
\end{equation}
Post-trade holdings are therefore
\begin{equation}
\begin{aligned}
w_{1}^{*}=\frac{\epsilon_1}{3}\,\mathcal{W}_6,\,\,\,
w_{2}^{*}=\frac{1-\epsilon_1}{3}\,\mathcal{W}_6,\\
w_{3}^{*}=\frac{\epsilon_2}{3}\,\mathcal{W}_6,\,\,\,
w_{4}^{*}=\frac{1-\epsilon_2}{3}\,\mathcal{W}_6,\\
w_{5}^{*}=\frac{\epsilon_3}{3}\,\mathcal{W}_6,\,\,\,
w_{6}^{*}=\frac{1-\epsilon_3}{3}\,\mathcal{W}_6.
\end{aligned}
\label{eq:k11}
\end{equation}

Note that these two considerations for hexadic exchange facilitate the check of the robustness of our 3-agent interaction model. 
For the first type of interaction, the random numbers are chosen following Eq.~\ref{eq:k2}, whereas for the other interaction type, the random numbers are drawn following the DY model, in addition to the fact that the total wealth is distributed evenly among the three pairs.

\subsection{Summary of exchange kernels}\label{sec:2.7}
Table~\ref{tab:1} synthesises the normalization adopted for each group size.
\begin{table}[H]
    \centering
    \begin{tabular}{|c|c|c|c|}
    \hline
       $\textbf{$n$}$  & \textbf{Constraint on raw uniforms} & \textbf{Scaling to obtain $\pi_i$} & \textbf{$\sum_i\pi_i$} \\
       \hline
    2 & $\epsilon\sim U(0,1)$ & $\pi_{1}=\epsilon,\ \pi_{2}=1-\epsilon$ & 1   \\
    \hline
    3  & $\sum\epsilon_i=\tfrac32$ & $\pi_i=\tfrac{2}{3}\epsilon_i$ & 1 \\ 
    \hline
4  & pairs, each half of wealth  & $\pi_{1,2}=\frac{\epsilon_1,\, 1-\epsilon_1}{2},\,\pi_{3,4}=\frac{\epsilon_2,\, 1-\epsilon_2}{2}$& 1 \\ 
\hline
5 & $\sum\tilde{\epsilon}_i=\tfrac52$ & $\pi_i=\tfrac{2}{5}\tilde{\epsilon}_i$ & 1 \\
\hline
6-I & two triads, each $\tfrac32$ & $\pi_i=\epsilon_i/3$ & 1 \\
\hline
6-II & three dyads, each $\tfrac13$ & see Eq.(\ref{eq:k10}) & 1 \\
\hline
    \end{tabular}
    \caption{Summary of exchange kernels.}
    \label{tab:1}
\end{table}

All rules satisfy conservation, symmetry, and uniform marginals, allowing us to isolate the sole effect of interaction arity on the emergent wealth distribution, analyzed in the following section.

\section{Simulation parameter choices and validation of the robustness}\label{sec:3}

\begin{figure}[tbh]
    \centering
\includegraphics[width=0.47\columnwidth]{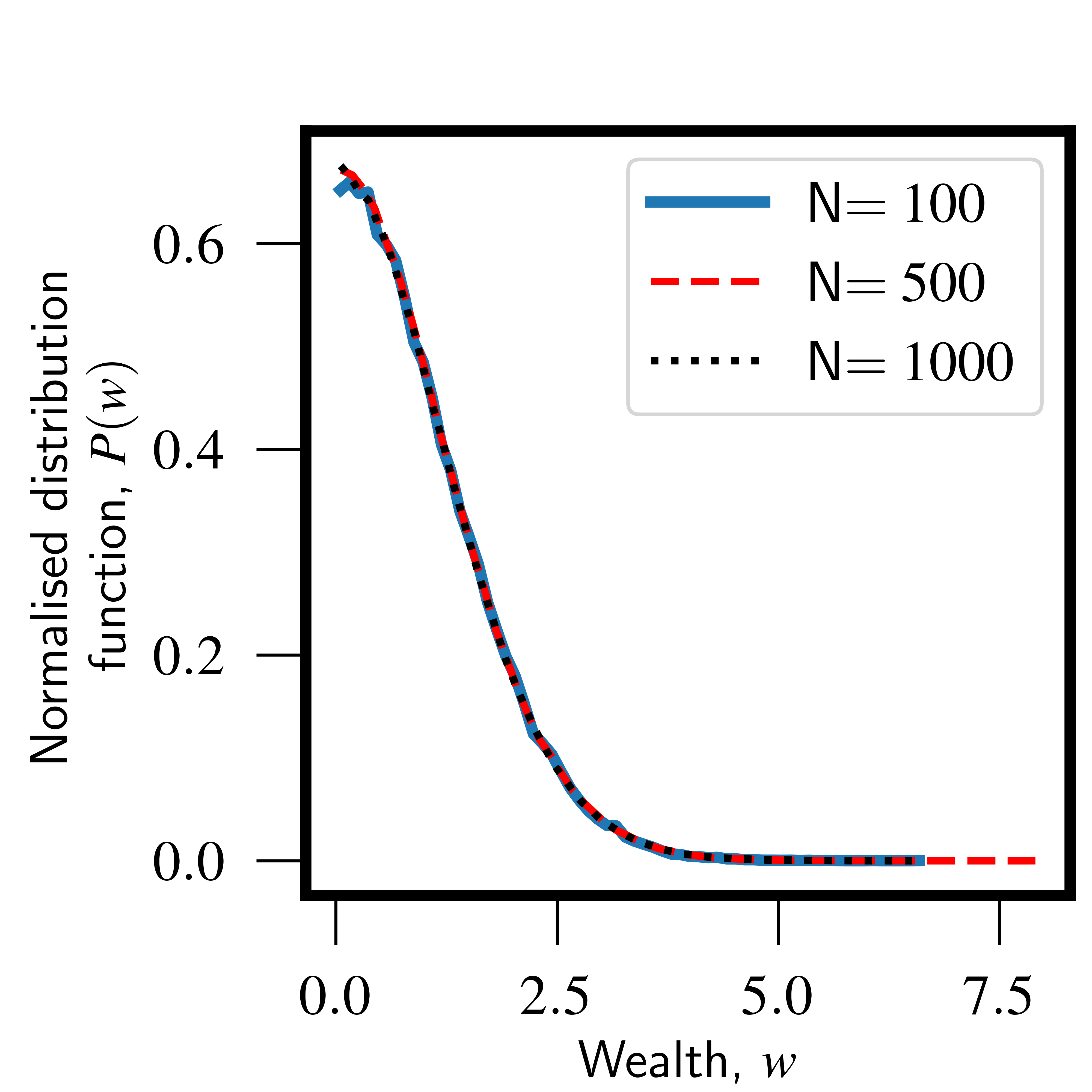}
\includegraphics[width=0.47\columnwidth]{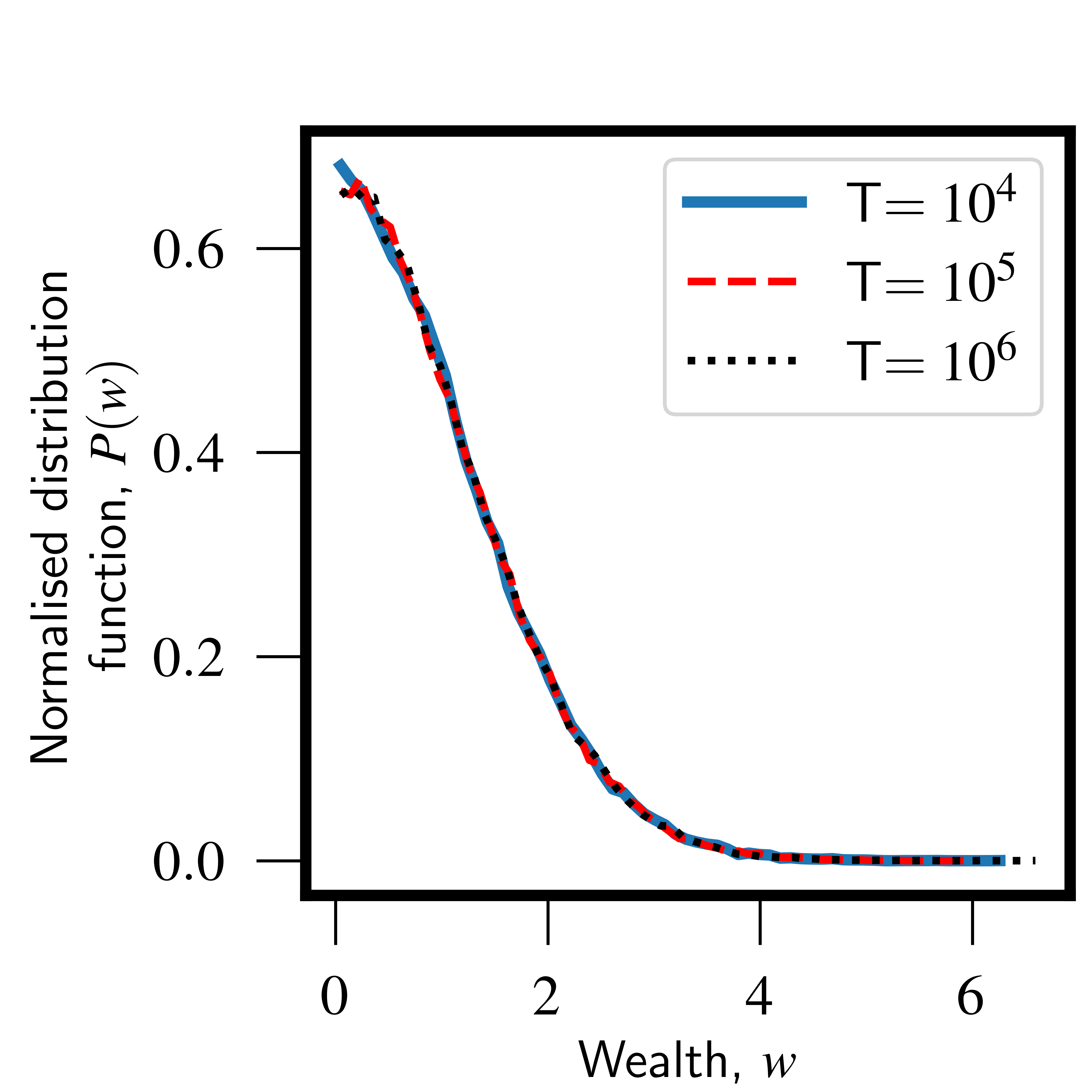}
\caption{Figure demonstrating the robustness checks for the triadic kernel. Left: collapse of probability density functions for different total number of agents (N) implying a steady state independent of number of agent; Right: overlap of the same for different number of total iteration (T) implying the convergence to a steady state.}
   \label{fig:size}
\end{figure}
We perform Monte Carlo simulations of wealth exchange dynamics following the rules specified in the earlier section.
For all the simulations presented here, we consider $10^{4}$ independent realizations of the system and perform an ensemble averaging to compute the wealth distribution.
Unless otherwise stated, all agents are initialized with unit wealth, so that the initial total wealth is equal to the total number of agents.

Further, to understand the steady behavior of the wealth distribution and the impact of the total number of agents, we perform two sets of numerical experiments with triadic interactions.
In the first experiment, we vary the total number of agents, $N \in \{100,500,1000\}$, while keeping the total number of collisions fixed at $T = 10^{6}$.
In the second experiment, we fix the number of agents at $N = 100$ and vary the total number of collisions, $T \in \{10^4, 10^5, 10^6\}$. 
For each choice of $(N,T)$, we compute the wealth distribution after the prescribed number of collisions and then average over $10^4$ realizations.
The results of the experiments are shown in Figure~\ref{fig:size}.
In the left panel of the figure, the distributions are shown after $10^{6}$ iterations and averaging over $10^{4}$ independent realizations for different numbers of total agents. 
The distribution function is plotted after a normalization with the mean wealth $\langle w\rangle=\text{Total wealth}/\text{Total number of agents}$.
We observe a complete overlap of the distribution functions within statistical uncertainty, implying its invariance on the system size. 
In the right panel, we present the normalized distribution functions with $N=100$ and for different numbers of total iterations. 
Here, we also observe a complete overlap of the distribution functions, implying the system's steady behavior following the triadic interaction. 
{The complete overlap of the distributions across different choices of $N$ and $T$ confirms that our subsequent results are robust with respect to finite size and finite collision time effects. Consequently, for all simulations presented in the remainder of the paper, we fix the system size to $N=100$ (with unit initial wealth per agent) and the total number of collisions to $T=10^{6}$.}

\begin{figure}[tbh]
    \centering
    \includegraphics[width=0.52\columnwidth]{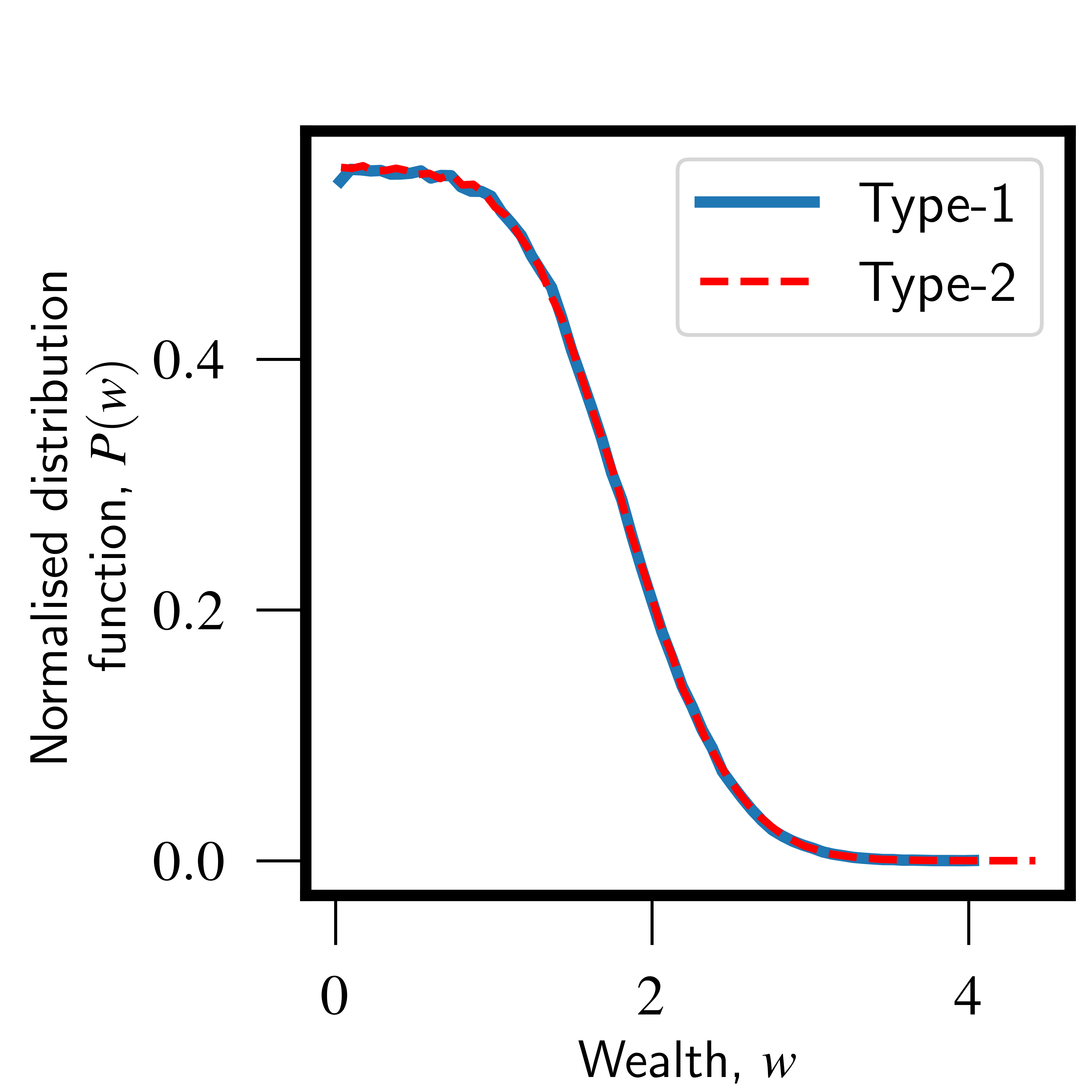}
    \caption{Steady state wealth distribution for 6-agent interaction considering two different types of interaction kernels (Eq.~\ref{eq:k9} and Eq.~\ref{eq:k11} respectively). This demonstrate the equivalence between the dyadic and triadic interactions.}
    \label{fig:topology}
\end{figure}

As a check of the robustness of the triadic interaction, we present in Figure~\ref{fig:topology}, the distribution functions obtained from the two different types of interactions, as mentioned earlier, for 6-agents.
One type considers a`two-triplet' kernel (Eq. \ref{eq:k9}) and the other type considers a `three-dyad' kernel (Eq. \ref{eq:k11}).
We observe an overlap of the distribution functions for both types, which implies the robustness of the triadic interaction against the known DY model of binary exchange. 

To quantify the statistical uncertainty in the numerically obtained distributions, we treat the $10^4$ realizations as independent samples. For a given choice of $(N=100,\,T=10^6)$, we construct a histogram of wealth using a fixed set of bins. Denoting by $p_{k}^{(r)}$ the normalized height of bin $k$ in realization $r$, the ensemble averaged distribution in bin $k$ is,
\begin{equation}
    \overline{p}_{k} = \frac{1}{R} \sum_{r=1}^R p_{k}^{(r)},\,\,\,\,\,\,\,\,\,\,\,\,R=10^4.
\end{equation}
The corresponding sample variance across realizations is,
\begin{equation}
    \sigma_{k}^2 = \frac{1}{R-1} \sum_{r=1}^R (p_{k}^{(r)}-\overline{p}_{k})^2,
\end{equation}
and the standard error of the mean in bin $k$ is $\sigma_{k}/\sqrt{R}$. Assuming approximate normality (justified by the large number of realizations), we construct a $95\%$ confidence interval in each bin as,
\begin{equation}
    \overline{p}_{k}\pm 1.96\frac{\sigma_{k}}{\sqrt{R}}.
\end{equation}
In Figure~\ref{fig:multi}, these $95\%$ confidence intervals are represented as shaded bands around the mean distribution on the binned data.

\section{Result and Analysis}\label{sec:4}

\subsection{Emergent wealth distributions for multi-agent exchange:}
\begin{figure}[tbh]
    \centering
    \includegraphics[scale=0.9]{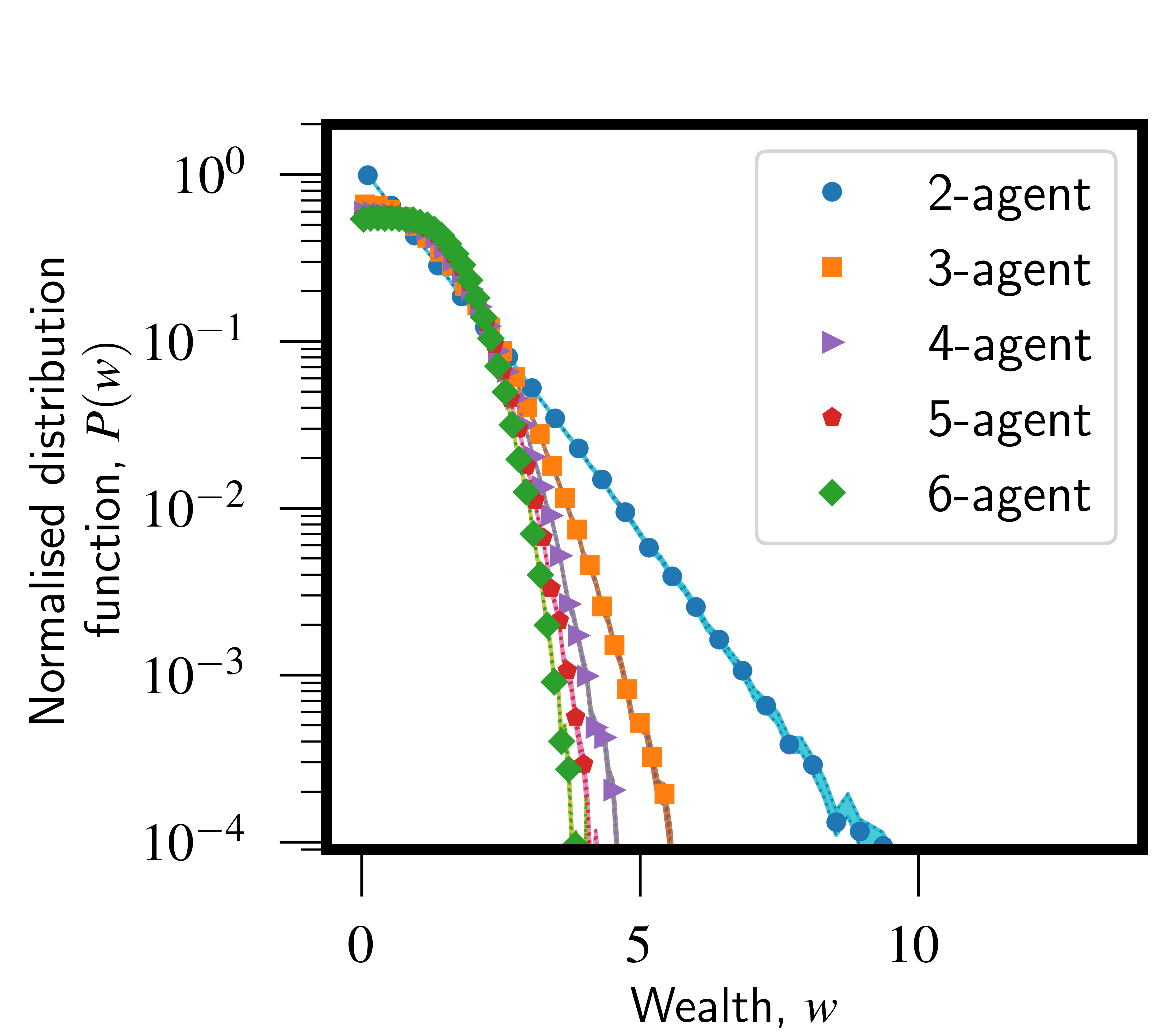}
    \caption{Steady state distribution of wealth for 3 agent, 4 agent, 5 agent and 6 agent wealth exchange models and comparison with 2 agent exchange model. Each curve showing $95\%$ confidence interval through the shaded regions.}
    \label{fig:multi}
\end{figure}

Figure~\ref{fig:multi} presents the stationary wealth distributions obtained after $T=10^6$ collisions for number of interacting agents $n=2,3,4,5,6$ (all curves are ensemble–averaged over $R=10^4$ realisations and normalised by the mean wealth $<w>\,=1$).
The 2-agent wealth distribution resembles the classical Boltzmann–Gibbs exponential $P(w)\propto e^{-w}$, which is shown here for comparison purposes. 
As the agent number is increased, we observe flattening of the distribution in the lower wealth region and a steeper fall in the higher wealth portion. 
The steepness increases with the number of agents, which implies that with the increase in exchange hands, the existence of riches ceases in the society.
Such an effect drives the inequality to reduce as compared to the 2-agent system. 
We also observe that, with the increase in the number of agents, the distribution function converges to a uniform wealth distribution. 
These observations establish the number of exchange hands as a relevant control parameter in understanding the wealth distribution in a society: larger groups redistribute money more evenly, even though the micro-rule remains purely random and conservative.

\subsection{Inequality metrics: Lorenz, Gini and Kolkata indices for multi-agent exchange:}

A central tool for assessing income and wealth disparities is the Lorenz function and its graphical representation is the Lorenz curve~\cite{lor,lor1}. 
The Lorenz curve (L(p)) shows the cumulative share of total income earned by cumulative fractions of the population, after ordering individuals from lowest to highest income. 
It is drawn within the unit square, beginning at $(0, 0)$ and ending at $(1, 1)$. 
The $45-$degree line represents perfect equality, where every individual receives the same income i,e, $L(p)=p$. In Fig.~\ref{fig:ineq}, we show a schematic diagram of Lorenz curve and some associated parameters.

To quantify the inequality from the Lorenz curve, there are several statistical measures present. 
Gini index\footnote{Corrado Gini introduced the Gini coefficient in 1912 as a statistical measure of concentration. He later authored political writings, including “The Scientific Basis of Fascism” (1927)~\cite{Gini1927Fascism}, and has been discussed in the historiography of statistics under Italian fascism. The index itself is a value‑free statistical functional widely used across disciplines; our use is purely methodological.}~\cite{gini}, one of the most celebrated inequality measure, is the ratio of the area between the equality line and the Lorenz curve to the total area under the equality line. Equivalently, the Gini index is twice the area between the Lorenz curve and the line of perfect equality. The value of gini, $g\in[0,1]$ and it is an area-based measure. So mathematically we can write, $$g=1-2\int_{0}^1L(p)dp=2\int_0^1(p-L(p))dp\,.$$

\begin{figure}[tbh]
\centering
\begin{tikzpicture}[scale=7]
\def\xp{0.618}                        
\def\yp{0.382}
\draw[thick] (0,0) rectangle (1,1);
\draw[thick,green!80!black] (0,0) -- (1,0);
\draw[thick,green!80!black] (1,0) -- (1,1);

\draw[fill=yellow] (0,0) -- plot[domain=0:1,samples=200,smooth] (\x,\x*\x) -- (1,1);
\draw[fill=pink] plot[domain=0:1,samples=200,smooth] (\x,\x*\x) -- (1,1) -- (1,0) -- (0,0);


\draw[thick,blue] (0,0) -- (1,1);

\draw[thick,red]plot[domain=0:1,samples=200,smooth] (\x,\x*\x);

 \draw[thick,green] (0,0) -- (1,0) -- (1,1);

\draw[thick,orange] plot[domain=0:1,samples=200,smooth] (\x,{(1-\x*\x)});

\draw[thin] (0,1) -- (1,0);

\coordinate (Q) at (\xp,0);
\draw[dashed] (Q) -- (\xp,\xp); 

\coordinate (O) at (0.5,0.5);
\coordinate (P) at (\xp,\yp);

\node[left]        at (O) {$O$};
\node[right]       at (P) {$P$};
\node[below]       at (Q) {$k$};

\node[above,black]  at (0,1) {$D\,\,(0,1)$};
\node[below,black]  at (0,0) {$A\,\,(0,0)$};
\node[above,black] at (1,1) {$C\,\,(1,1)$};
\node[below,black] at (1,0) {$B\,\,(1,0)$};

\node[rotate=90] at (-0.08,0.5)
  {Cumulative fraction of wealth};

\node at (0.5,-0.12)
  {Cumulative fraction of population};

\node[right] at (1.05,0.80) {perfect equality line};
\draw[->] (1.045,0.80) -- (0.80,0.80);

\node[right] at (1.05,0.63) {Lorenz curve};
\draw[->] (1.045,0.63) -- (0.79,0.63);

\node[right] at (1.05,0.40) {perfect inequality line};
\draw[->] (1.045,0.40) -- (0.99,0.40);

\node[right] at (1.05,0.18) {complementary Lorenz curve};
\draw[->] (1.045,0.18) -- (0.90,0.18);
\end{tikzpicture}
\caption{Schematic diagram of Lorenz curve and related parameters. Gini index, $g$ is the ratio of the area between the yellow region and (yellow$+$pink) region. Kolkata index is the ordinate point, $k$ where Lorenz curve cuts the off diagonal line $\overline{BOPD}$.}
\label{fig:ineq}
\end{figure}

Another inequality index is the Kolkata index, or $k$-index, which was introduced in Ref.~\cite{k-index} and subsequently developed and applied in Refs.~\cite{k-index1,k-index2,k-index3,sand}. The $k$-index measures the share of wealth possessed by the richest $(1-k)$ fraction of the population, with $k\in[1/2,1]$.
This index has been used in literature, along with the Gini index, to characterize inequality and phase transitions in a range of physical and socio-economic systems~\cite{sand,Phys}. 
We adopt this index in addition to the Gini, to understand their functional dependencies for multi-agent interactions. 
There are several properties that this index bears, and some of them are given in what follows. 
This is a unique non-trivial fixed point of the complementary Lorenz curve, which directly gives the tail share. So mathematically $k$-index is the unique solution of $L(k)=1-k$, so that the richest $(1-k)$ fraction of agents hold exactly the top $k$ fraction of wealth.
While $g$ averages inequality across the entire distribution (area under the Lorenz curve), sensitive to changes across the distribution; whereas $k$ focuses on the top share vs rest through a fixed point, more interpretable for how the top tail is concentrated. 
Initially they start with different values, eventually it has been observed that they coincide to a value (nearly equal to $0.87$) below the extreme inequality value, $1$. 
Empirically, for many distributions with moderate inequality one often observes an approximately linear relation $k\approx\frac{1}{2}+\gamma g$ with $\gamma\approx0.37$, then $g$ and $k$ coincide at a value afterthat their relation become nonlinear upto the extreme value where they again coincide. The value of $\gamma$ is not a universal identity; deviations appear with heavier tails or pronounced middle‑mass. Before the coincidence point $k$ values are greater that $g$ as obviously $k$ starts with higher value but after the coincidence point the value of $g$ become greater than $k$ and they meet again at $1$ for a distribution.

\begin{figure}[tbh]
    \centering
    \includegraphics[width=0.49\columnwidth]{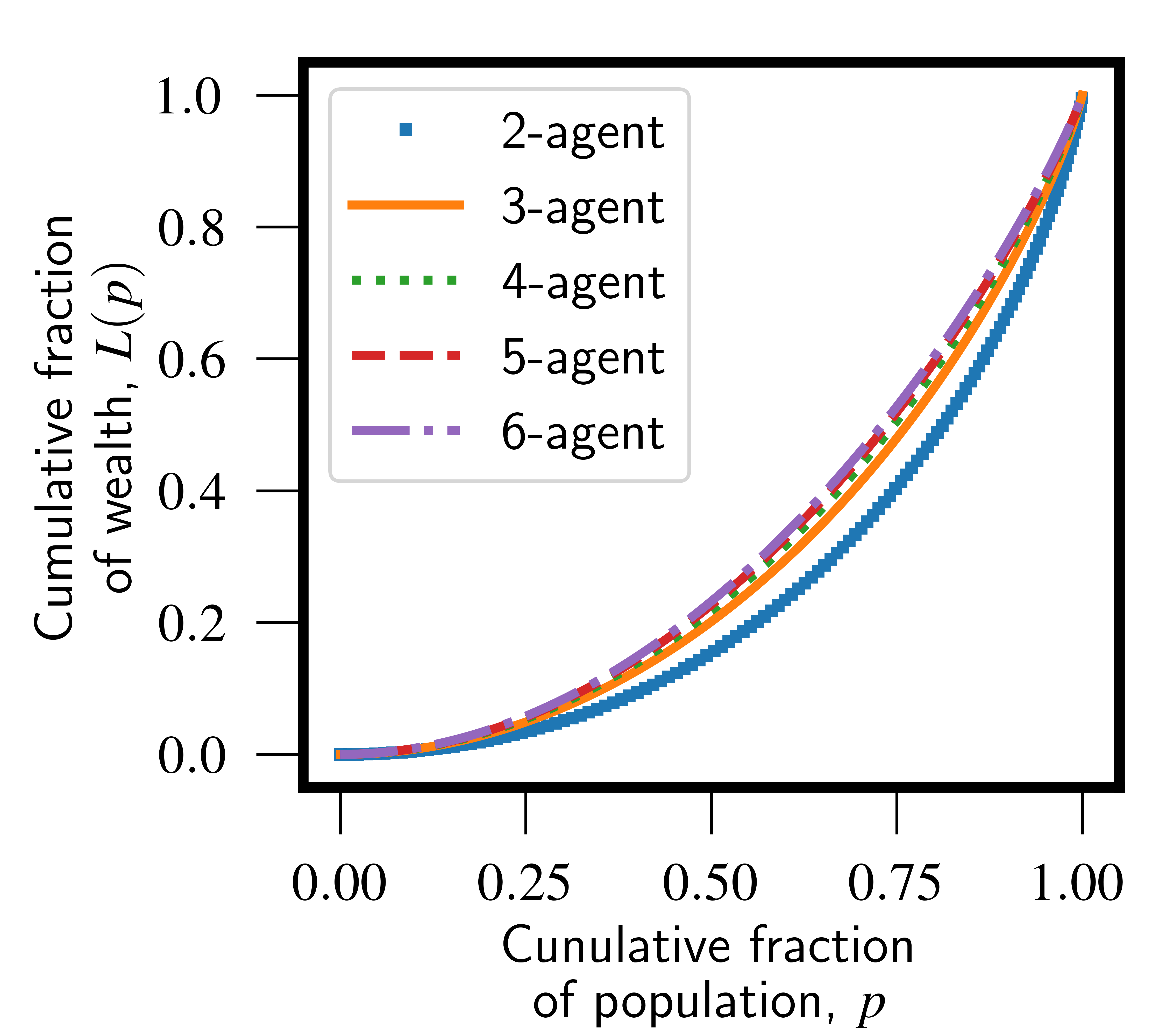}
    \includegraphics[width=0.49\columnwidth]{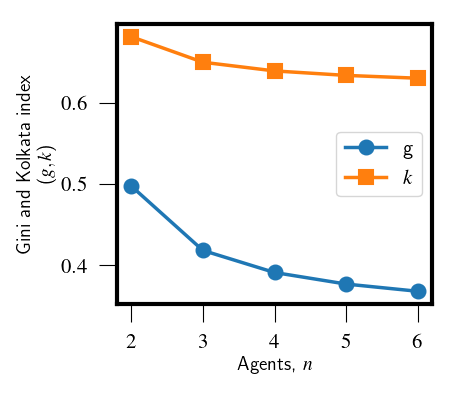}
    \caption{Figures depicting (left) Lorenz curves; (right) Gini and Kolkata indices for different multi-agent interactions.}
    \label{fig:lor}
\end{figure}

The left panel of Figure~\ref{fig:lor} presents Lorenz curves~\cite{lor,lor1} for all five kernels described in Table~\ref{tab:1}. 
Each curve bows progressively closer to the egalitarian diagonal as $n$ increases, confirming the decrease in inequality with the increase in the number of exchange hands.
Table~\ref{tab:2} quantifies the inequality matrices for different numbers of agents.
One observes a monotonic decrease in the inequality measures as $n$ increases, which is further demonstrated in the right panel of Figure~\ref{fig:lor}.
\begin{table}[H]
    \centering
    \begin{tabular}{|c|c|c|}
    \hline
 No. of agents & Gini ($g$) & $k$-index \\
\hline
 2  & 0.497 $\pm$ 0.004 & 0.681 $\pm$ 0.006 \\
 3   & 0.418 $\pm$ 0.003 & 0.650 $\pm$ 0.005 \\
 4  & 0.391 $\pm$ 0.003 & 0.639 $\pm$ 0.004 \\
 5  & 0.377 $\pm$ 0.002 & 0.634 $\pm$ 0.003 \\
 6  & 0.368 $\pm$ 0.002 & 0.630 $\pm$ 0.003 \\
\hline
    \end{tabular}
    \caption{Numerical values of gini and kolkata indices for different multi-agent interaction kernels.}
    \label{tab:2}
\end{table} 
 The Kolkata index or $k$-index~\cite{k-index,k-index1,k-index2,k-index3}, approaches the theoretical lower bound $0.63$ expected for the perfectly uniform distribution limit. Similarly, the standard inequality measure, Gini ($g$)~\cite{gini} becomes $0.36$, which is also close to the theoretical value of $g$ as observed for uniform distribution. So from the Table~\ref{tab:2}, we expect that as the number of agents increases, the wealth distribution approaches from exponential distribution towards uniform distribution. 

\begin{figure}[h]
    \centering
    \includegraphics[width=0.6\columnwidth]{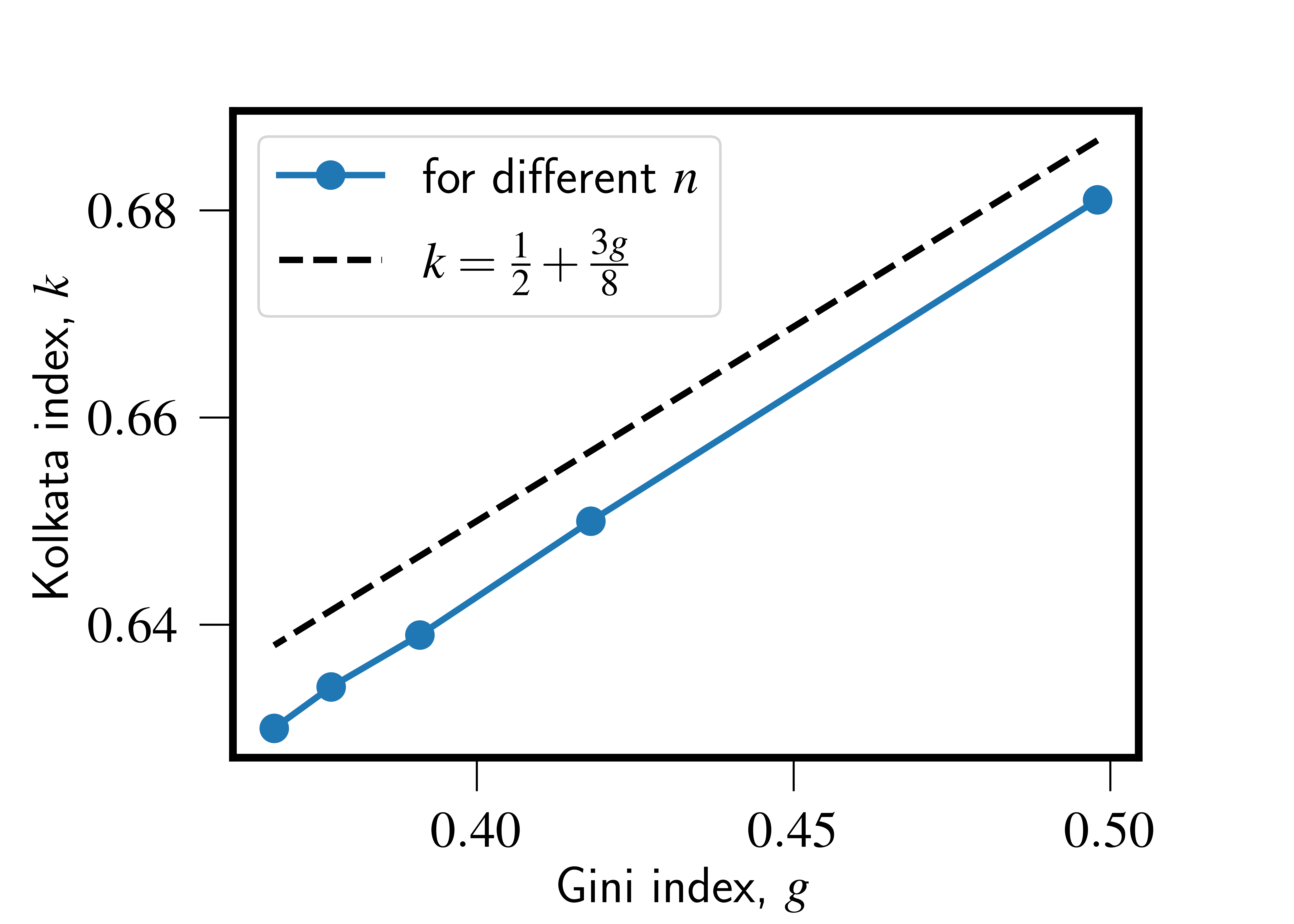}
    \caption{Figure demonstrating a comparison between gini and kolkata indices for different multi-agent interactions.}
    \label{fig:uni}
\end{figure}

Figure~\ref{fig:uni} presents $k$-index against Gini for the five cases. 
All points fall close to the empirical line $k=0.5+0.375g$ (black dashed), observed earlier in real-world income data and analytically derived for Gamma-like pdfs~\cite{sand,bijin}. 
{As in this model the maximum inequality is obtained when binary interaction happens and we get the exponential wealth distribution. If $n$ increases, we can see here that inequality decreases, so for this model we only get the initial almost linear portion of the $g$-$k$ relationship upto the values for the exponential distribution one. 
Such a linear relationship between Gini and k index validates the observations made in earlier literature.}

\subsection{Comparison of multi-agent exchange with binary saving-propensity models}
\begin{figure}[h]
    \centering
    \includegraphics[width=0.7\columnwidth]{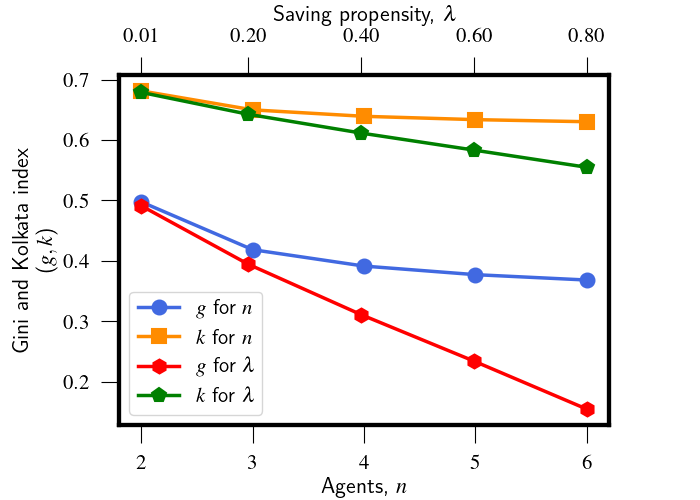}
    \caption{Figure showing variation of gini and kolkata indices for different multi-agents interaction kernels. A comparison based on binary kinetic wealth exchange including saving propensities is shown.}
    \label{fig:comp}
\end{figure}

As mentioned earlier, the triadic interaction seemed like some amount of wealth is being reserved in the agents hand and the rest is getting redistributed. 
Therefore we compare the behavior of the decrease in inequality due to the increase of the number of exchange hands with the binary DY like exchange with savings propensity, as such kind of interactions are based on the principle that the proportional wealth to be retained in hand and the rest is redistributed. 
We present the result in Figure~\ref{fig:comp}.
We observe that both parameters (saving propensity $\lambda$ and number of agents $n$) lower inequality as expected, but at different rates.
Increase in the saving propensity drives the inequality to fall monotonically, leading to a more and more egalitarian society, whereas increase in the number of agents shows more of a logarithmic decrease, converging to a non zero value as $n$ increases. 
High saving propensities are macroscopically egalitarian but micro-economically illiquid: circulation slows because only a shrinking fraction $(1-\lambda)$ is ever put at stake. 
In real economies, it is neither feasible nor desirable to drive $\lambda$ too close to unity—hoarding starves investment and growth. 
Raising the number of interacting hands avoids that trade-off: the entire group wealth is reshuffled at each collision, maintaining full turnover while still dampening the inequality.
Coordinating larger-than-binary transactions (crowd-funded loans, group procurement, employee ownership schemes) is typically easier than legislating universal high-saving behavior. 
The Figure~\ref{fig:comp} therefore highlights a practical window: modest increases in group size $(n \sim 4-6)$ deliver a sizable share of the inequality reduction achieved by money circulation, yet keep markets liquid.

\section{Discussion}\label{sec:5}

In this work we have shown that allowing money to circulate among more agents in a single interaction systematically reduces inequality in a kinetic-exchange setting. As soon as exchanges involve more than two agents, the stationary wealth distribution departs from the standard exponential (Boltzmann–Gibbs) form that characterizes the reversible Dragulescu--Yakovenko (DY) binary model. This outcome is qualitatively intuitive, yet it stands in contrast to rigorous results by Lanchier~\cite{lachier2017} and Makarov~\cite{JPC2025}, where reversible $n$-agent redistribution kernels on the simplex yield exponential steady states irrespective of $n$. Our findings therefore highlight the role of non-reversible, higher-order interactions as a distinct mechanism for generating non-Boltzmann wealth distributions.

The inequality reduction we observe as interaction arity increases is reminiscent of models with saving propensity, where a fraction of wealth is systematically retained by agents and thus not fully exchanged. However, a direct comparison with saving-based dynamics shows that the pattern of inequality reduction in our model is different, indicating that multi-agent pooling and redistribution behaves fundamentally differently from saving-propensity–driven mechanisms. In our case, the key ingredient is not wealth retention but the way pooled wealth is redistributed across more than two agents in a single, non-factorizable event.

Methodologically, our focus is to isolate a single axis of complexity—\emph{interaction arity under global pooling}—while controlling the reshuffling noise as tightly as possible. For this reason we use the DY model as a minimal baseline with a well-understood reversible binary limit and a canonical exponential steady state. Within this baseline, deviations from the Boltzmann–Gibbs form under our multi-agent pooling kernels can be attributed directly to changes in interaction arity and pooling architecture, rather than to additional ingredients such as wealth-at-stake rules, multiplicative risk, or condensation mechanisms that are characteristic of yard-sale–type models. Our contribution should therefore be viewed as a controlled multi-agent generalization of kinetic exchange, rather than as a claim of maximal microeconomic realism. Extending the same pooling and arity architecture to yard-sale–type rules, and studying the combined effect of higher-order interactions and wealth-at-stake dynamics, is a natural direction for future work.

From a modeling standpoint, we do not attempt to construct the most general $n$-body redistribution kernel on the $(n\!-\!1)$-simplex. Instead, we introduce a restricted but transparent family of \emph{multi-agent pooling kernels} built from dyadic and triadic building blocks. For a selected group of $n$ agents we first pool their wealth, $W = \sum_{i=1}^n w_i$, choose integers $a,b \ge 0$ such that $n = 2a + 3b$, split the pool into $a+b$ equal pots of size $W/(a+b)$, and then apply the binary DY reshuffling rule to each dyadic pot and our triadic reshuffling rule to each triadic pot. This global pooling and equal-pot architecture is not equivalent to simultaneous independent binary DY: in the latter, each pair redistributes its own pre-trade pair sum and the $n$-body transition kernel factorizes into a product of two-body operators, whereas in our scheme the exchanged amounts are set by the global pool $W$ and the pot decomposition, so the $n$-body operator does not factorize and each agent’s update depends on the full $n$-agent pool.

Although real economic interactions are considerably more complex, the present framework captures a simple but robust qualitative effect: when wealth circulation occurs through genuinely multi-agent pooling events rather than purely pairwise trades, inequality is systematically reduced and the stationary distribution can move away from the Boltzmann–Gibbs benchmark. This is consistent with the fact that many real-world institutions rely on inherently multi-agent interactions, such as oligopolistic firms jointly choosing prices or quantities~\cite{cournot_1838,tirole_1988}, auctions and platform-mediated markets, financial clearing mechanisms, and various forms of group-based wealth management in families, coalitions, boards, or investment syndicates. Rather than providing a microfoundation for perfectly competitive equilibrium, the present model offers a minimal non-equilibrium extension of kinetic exchange that is capable of producing non-Boltzmann steady states, and thereby serves as a useful conceptual laboratory for studying how multi-agent market interactions can reshape inequality.

\section{Summary and Conclusion} \label{sec:6}
In this study, we examine the influence of multi-agent interaction dynamics on emergent wealth distributions within the framework of kinetic exchange models.
Departing from the classical binary exchange approach, we propose a new interaction rule involving simultaneous wealth exchange among three agents. 
This triadic rule is conceptualized as an extension of the standard kinetic wealth exchange model, which has traditionally focused on pairwise (binary) transactions.
With such extension we evaluate the multi-agent (more than 3) wealth interaction scenario as a mixture of binary and triadic interaction processes, allowing for a richer and more interconnected model of economic exchange.

Through detailed numerical simulations, we analyze how the inclusion of higher-order interactions alters the statistical properties of wealth distribution in a closed economy. 
Our findings reveal that increasing the number of agents participating in a single exchange event significantly modifies the wealth distribution profile. 
Compared to the canonical exponential distribution produced by binary kinetic models, the multi-agent model yields a distribution that is flatter in the lower-wealth region and exhibits a sharper cutoff in the high-wealth tail. 
This indicates a redistribution of wealth from the extremes toward more moderate levels, suggesting a dampening of both poverty and extreme wealth accumulation.

We assess inequality using two standard indices: the Gini coefficient and the $k$-index, the latter being a recently proposed measure that captures inequality based on the minimum $k$ fraction of the population holding a maximum $(1-k)$ fraction of wealth. 
Our results show that both indices decline as the prevalence of multi-agent interactions increases, indicating a general reduction in inequality. 
This observation supports the hypothesis that broader participation in economic exchanges facilitates more equitable wealth circulation.

To contextualize our model, we compare the inequality dynamics in our multi-agent framework with those in the binary exchange model incorporating saving propensity—a well-established mechanism for reducing inequality. 
While the saving-propensity model leads to a monotonic reduction in inequality, ultimately driving it to zero in the limit of maximum saving, the multi-agent model exhibits a qualitatively different behavior: inequality decreases more gradually and asymptotically approaches a nonzero value. 
This suggests that while multi-agent interactions promote greater equity, they may also preserve some inherent heterogeneity in the wealth distribution, reflecting real-world socioeconomic diversity.

From a broader perspective, the relevance of our findings extends to real economic systems where transactions often involve more than two participants. 
The incorporation of multi-agent interactions provides a more realistic modeling framework that captures the complexity of modern economic activity.
Furthermore, the observed relationship between the number of agents participating in interaction and inequality highlights potential pathways for policy interventions aimed at promoting inclusivity. 
Encouraging mechanisms that broaden economic participation—such as cooperatives, community-based finance, and decentralized trade platforms—may contribute to a more equitable distribution of resources without completely eliminating wealth differentials necessary for innovation and risk-taking.

\end{document}